\documentclass{proceedingsM}
\usepackage{times,amsmath,amssymb}
\usepackage{graphicx,color}
\usepackage[colorlinks=true, citecolor=blue]{hyperref}

\DeclareMathOperator*{\sech}{sech}
\title{Wave propagation in pantographic 2D lattices with internal discontinuities}
\author{Angela Madeo\footnotemark[1]$^,$\footnotemark[2],
Alessandro Della Corte\footnotemark[3], Leopoldo Greco\footnotemark[2]$^,$\footnotemark[4], Patrizio Neff\footnotemark[5]}
\address{{\footnotemark[1]} Universit\'{e} de Lyon-INSA (Institut National des Sciences Appliqu\'{e}es), Laboratoire de G\'{e}nie Civil et Ing\'{e}nierie Environnementale (LGCIE) B\^{a}timent Coulomb, 69100 Villeurbanne, France}
\address{{\footnotemark[2]} Dep. of Mechanical and Aerospace Engineering, Universit\`{a} di Roma La Sapienza, 18 Via Eudossiana, Roma}
\address{{\footnotemark[3]} MeMoCS, International Research Center for the Mathematics \& Mechanics of Complex Systems, Universit\`{a} dell'Aquila}
\address{{\footnotemark[4]} Department of Civil, Building-Architecture and Environmental Engineering, DICEA, University of L'Aquila}
\address{{\footnotemark[5]} University of Duisburg-Essen, Facoulty of Mathematics, Mathematik-Carr\'ee Thea-Leymann-Strasse 9 45127 Essen}
\email{angela.madeo@insa-lyon.fr, alessandro.dellacorte@uniroma1.it, leopoldo.greco@virgilio.it, patrizio.neff@uni-due.de}

\abstract{In the present paper we consider a 2D pantographic structure composed by two orthogonal families of Euler beams. Pantographic rectangular `long' waveguides are considered in which imposed boundary displacements can induce the onset of traveling (possibly non-linear) waves. We performed numerical simulations concerning a set of dynamically interesting cases. The system undergoes large rotations which may involve geometrical non-linearities, possibly opening the path to appealing phenomena such as propagation of solitary waves. Boundary conditions dramatically influence the transmission of the considered waves at discontinuity surfaces. The theoretical study of this kind of objects looks critical, as the concept of pantographic 2D sheets seems to have promising possible applications in a number of fields, e.g. acoustic filters, vascular prostheses and aeronautic/aerospace panels.}

\begin{document}
\maketitle

\section{Introduction}
\label{intro}

In the present paper, we employed the concept, suggested by prof. dell'Isola, of pantographic lattices, whose technological importance is rapidly increasing and, especially in nano-technology, could be very relevant. Pantographic structures are mechanical systems in which array of beams or rods are connected by internal kinematic pivots. Actually, a very wide class of objects can be effectively described and studied by means of suitably chosen pantographic models (see e.g.~\cite{dellIsola2014}). In this work, a numerical analysis of wave propagation is performed on the basis of the discrete mechanical model presented in~\cite{dell'Isola2015}. 

Let us briefly recall the model considered therein. In Fig. 1 (left), the reference configuration $C^*$ is shown. Lines indicate beams, which are divided in two families of parallel and equally spaced beams (with distance $d$), reciprocally orthogonal in $C^*$. The beams are arranged in a rectangle (sized $\sqrt{2}Ld \times \sqrt{2}Wd$ in $C^*$) whose sides are crossed by the beams at 45 degrees in $C^*$. 
Each beam has a standard linearized Euler strain energy given by:

\begin{equation}
\mathcal{E}=\int_{\Lambda} \frac{k_M(u'')^2+k_N(w')^2}{2}.
\end{equation}

Here $u$ and $w$ are respectively the values of transverse and axial displacements $\textbf{u}$ and $\textbf{w}$ with respect to~$C^*$, and $k_M$, $k_N$ are respectively bending and axial stiffness coefficients, which in the real object depend of course on the diameter of beams, while the integral is extended over the entire length $\Lambda$ of the beam.
    
Dots in Fig. 1 (left) represent hinges which allow free rotations and do not interrupt the continuity of the beams. The configuration at a given time $t$ can be thought as characterized by a `large' displacement (with respect to $C^*$) due to the contribution of the rigid rotations, and a `small' displacement due to axial and bending elastic deformations.

\begin{figure}
\begin{center}
\includegraphics[height=0.5\textwidth]{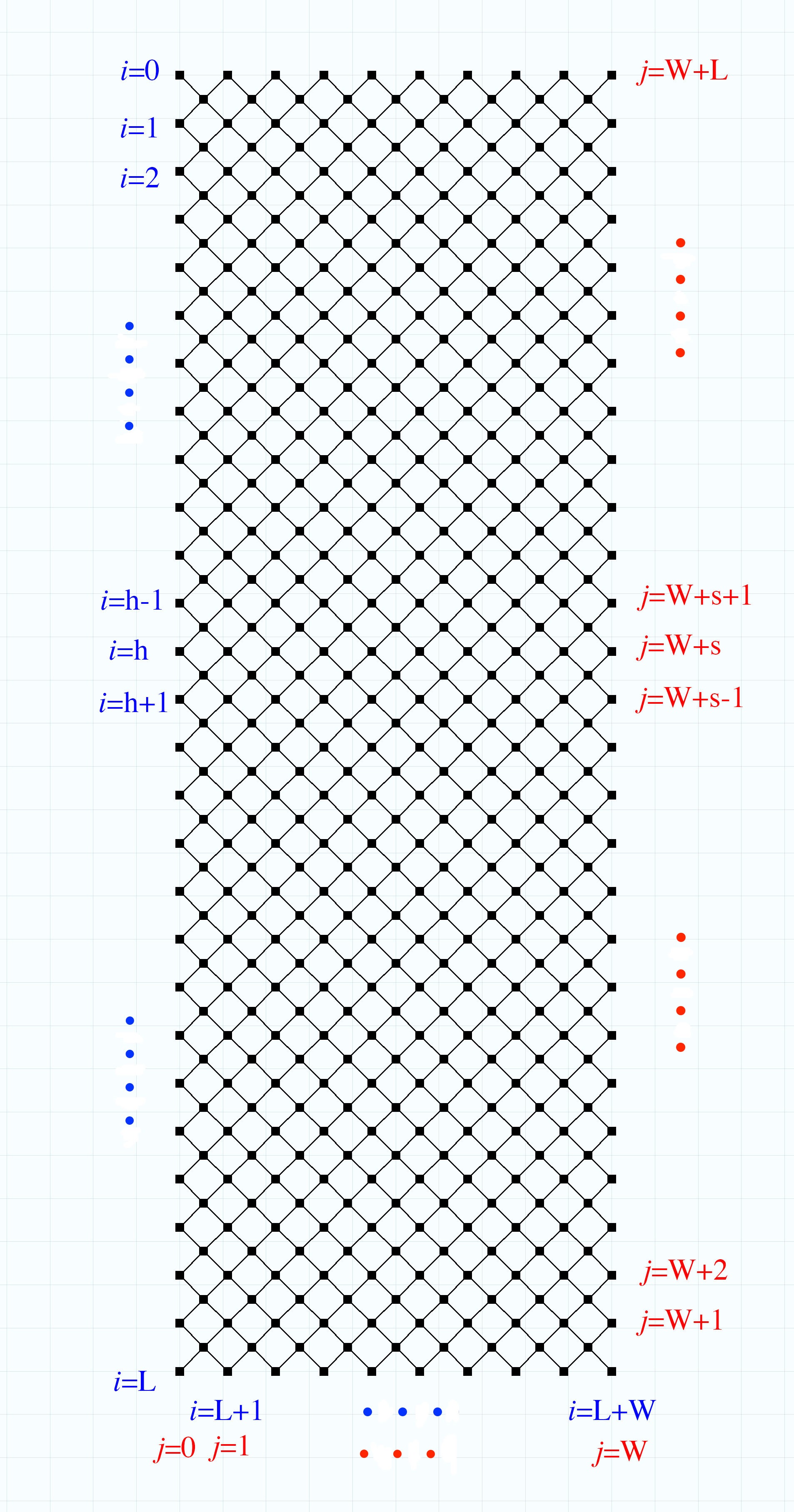} \quad \includegraphics[height=0.5\textwidth]{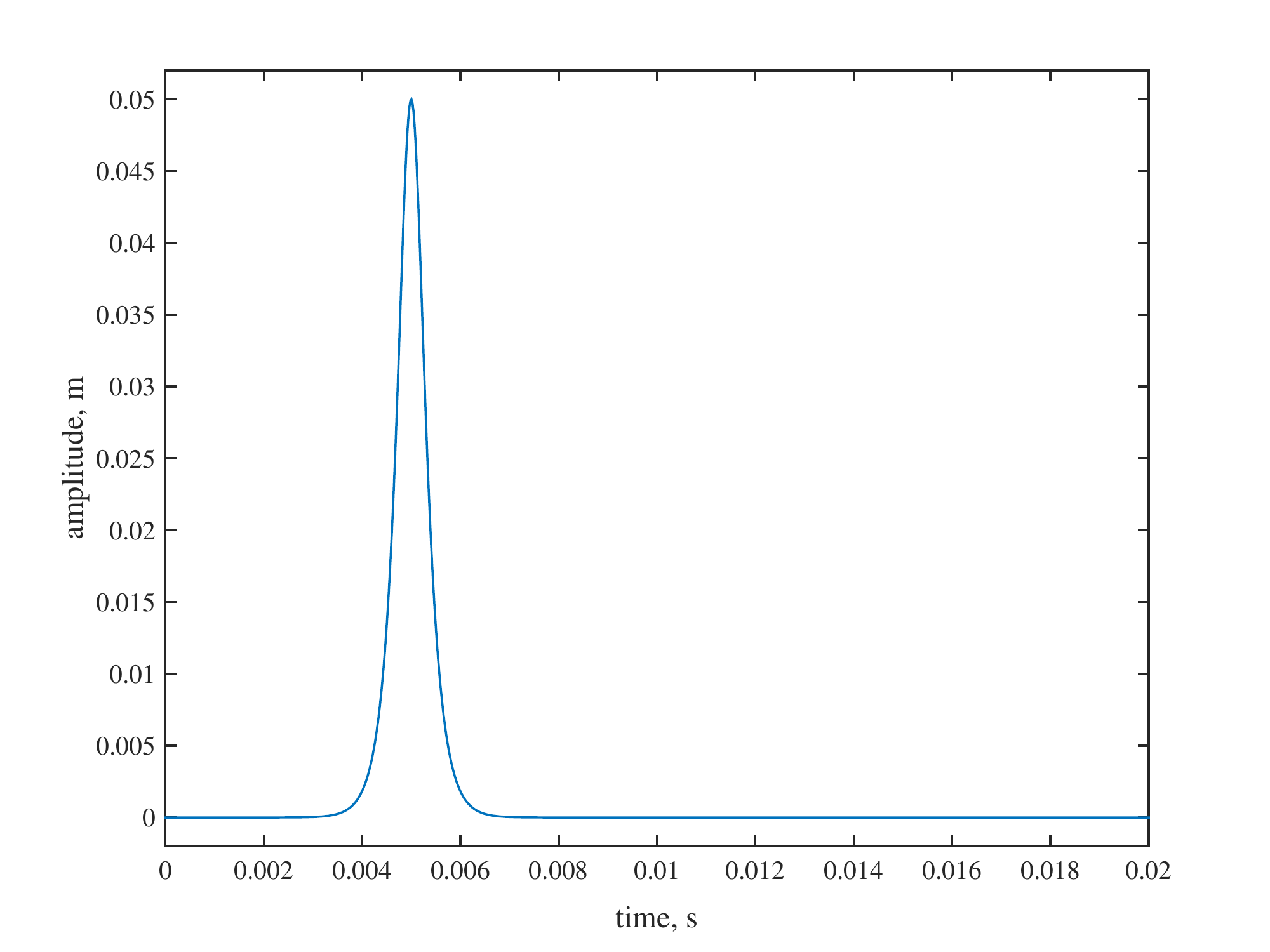}
\caption{Reference Configuration (left) and time history of the impulse (right)}
\end{center}
\end{figure}

\section{Numerical results}

Wave propagation in non-trivial structures are of course widely studied in literature (see e.g. ~\cite{Sestieri2003,Placidi2008,Quiligotti2002}), and in particular several numerical and experimental results on woven fabrics are found in literature (see e.g.~\cite{Boisse2008}).
For our numerical analysis we used a length $l$ of 0.1 m for the lower and upper sides, while the height $h$ of the rectangle is 2 m in simulations plotted in Figs 4 and 5 and 2.5 m in those plotted in Figs. 2, 3 and 6.

We chose the values $1.96 \times 10^{-2}$ N m$^2$ and $7.85 \times 10^4$ N for $k_M$ and $k_N$ respectively, which can be thought as relative to a beam with an elliptic section of semiaxes $a=0.001$ m and $b=0.00025$ m (area $A=7.85\times 10^{-7}$ m$^2$, inertia moment of the cross section around its minor axis $J=1.96\times 10^{-13}$ m$^4$) rotating around the minor one. We set $d=0.0(1)$ m for the simulations shown in Figs 3-4 and $d=0.00(5)$ m for the other ones.

As for mechanical parameters, we chose the mass density $\rho=1450$ kg/m$^3$, $Y=100$ GPa for the Young's Modulus and $\nu=0.2$ for Poisson's ratio. The material was assumed to be linearly elastic.

For all our simulations, we imposed a displacement on the points of the system belonging to the upper side of $C^*$, oriented along the height of the rectangle. The displacement is analytically represented by an impulse function $\mathfrak{I(t)}=u_0*\sech[\tau(t-t_0)]$, where $u_0=0.05$ m and $t_0=0.005$ s, while $\tau$ is a parameter affecting the duration of the impulse; in Fig. 1 (right) the impulse is plotted with $\tau=4000$ s$^{-1}$. In all the figures, times (in s) are given on the horizontal axis. If not otherwise specified, the lower side is built in, and the absolute values of the rotations of the cross sections are represented by a color map. 

The numerical problems that may arise when considering such kind of structures with peculiar geometric characteristics can be very complex, and moreover the behavior of the system can very easily display instabilities of the type of those discussed in, e.g., \cite{Luongo1998,Luongo2007,Luongo2005,Bersani2013,Piccardo2014}. A set of numerical tools has been elaborated to take care of this kind of problems (the reader is referred e.g. to \cite{Boisse2013,Cazzani2004,Cazzani2014,Cazzani2012,Cuomo2000}), also in direction of extending the model to the case of inextensible fibers, which can be numerically addressed by means of Lagrangian Multipliers Methods, as performed in (\cite{Cuomo1998,Cuomo2014}).

Of course, our numerical results are intended as a first step towards a general homogenized theory of this kind of systems. Homogenization problems of this kind can in fact be very difficult, but a series of related problems have already been attacked in the literature (see e.g. \cite{Yang2010,Yang2011,Seppecher2011,Lekszycki2012,Andreaus2014}). These homogenization methods are of course at the basis of many nowadays active lines of investigations, such as mechanical phase transition(~\cite{Eremeyev1991,Yeremeyev2003,Pietraszkiewicz2007,Eremeyev2009}), dissipation in particular structures (~\cite{Carcaterra2011}) and anisotropy problems (~\cite{Placidi2005}). 

Considering our pantographic system, it is reasonable to describe its homogenized limit as a metamaterial (see e.g. ~\cite{Del_Vescovo2014,Birsan2012,Andreaus2004,Maurini2004,Giorgio2009}), and since, as we will numerically show, the system responds to solicitations like double forces, higher-gradient theories are also called for (see e.g. ~\cite{Alibert2003,Isola2014,Javili2013,Rinaldi2014}). 

All numerical simulations are performed with \textsl{COMSOL Multiphysics}$^\text{\textregistered}$.

\subsection{Basic wave propagation and double impulse}
In Fig. 2 a basic case of wave propagation after an impulse of the type of Fig. 1 (right) is shown ($\tau=4000$ s$^{-1}$, height $h=2.5$ m). This results in quite ordinary (continuous-like) wave propagation. Dispersion is clearly observable, as the length of the perturbed zone is clearly increasing in time, and a reflection on the lower side is also visible in the last snapshots.

In Fig. 3, the effect of a double impulse applied in the middle height of the structure is shown ($\tau=4000$ s$^{-1}$, height $h=2.5$ m). By double impulse we mean a displacement oriented in the beam direction and imposed on two points belonging to the opposite ends of two adjacent beams - i.e. the ends are almost in line with one of the two families of beams. Both the upper and lower sides are built in. The idea is to show that such a structure can respond to a stimulus that, in the continuous homogenized limit case, is a double force (i.e. a pair of forces with null resultant and moment). Indeed, the onset of standard traveling waves is visible. 

\begin{figure}
\begin{center}
\includegraphics[width=0.6\textwidth]{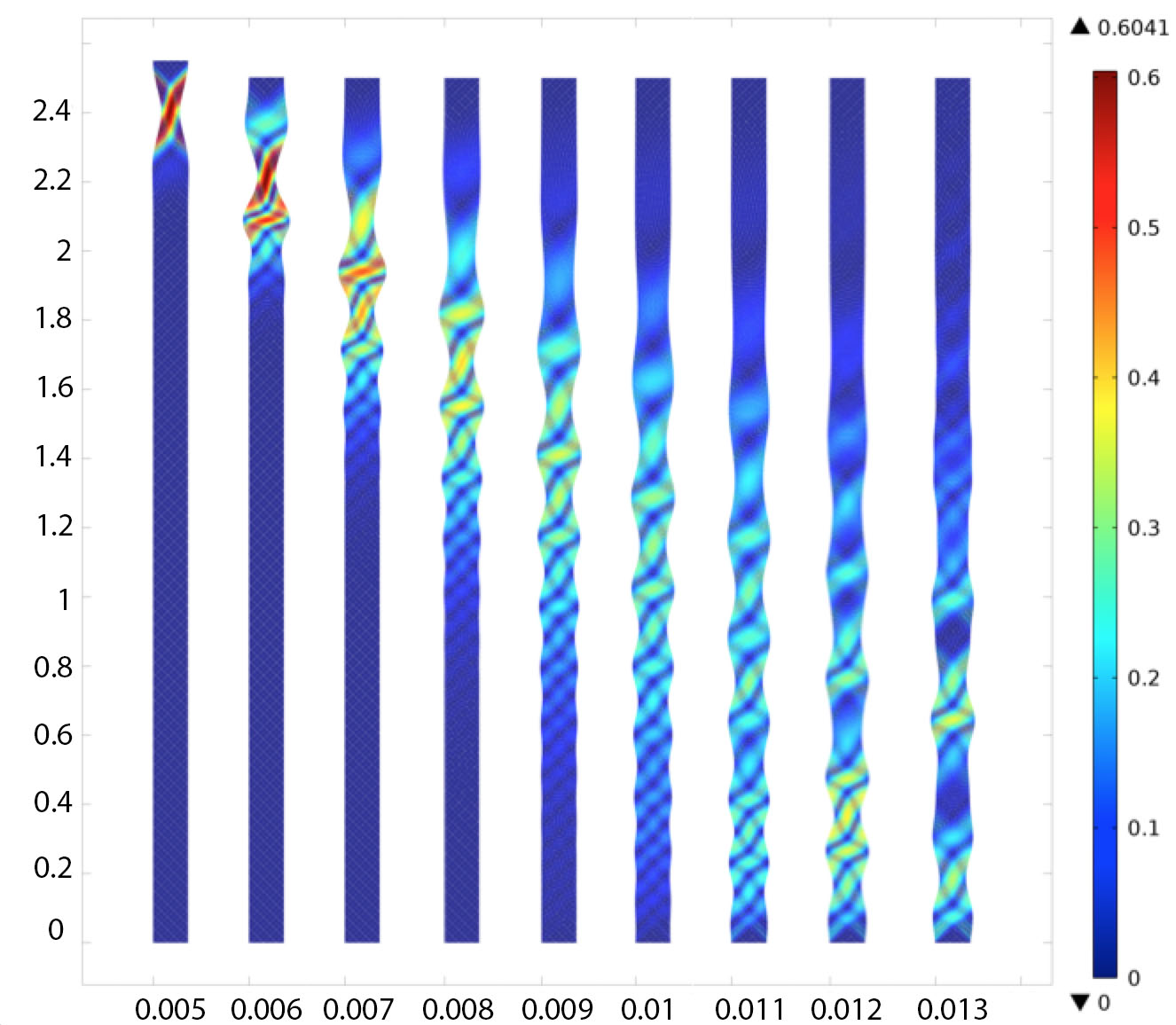}
\caption{Wave propagating after an imposed vertical displacement on the upper side}
\end{center}
\end{figure}

\begin{figure}
\begin{center}
\includegraphics[width=0.6\textwidth]{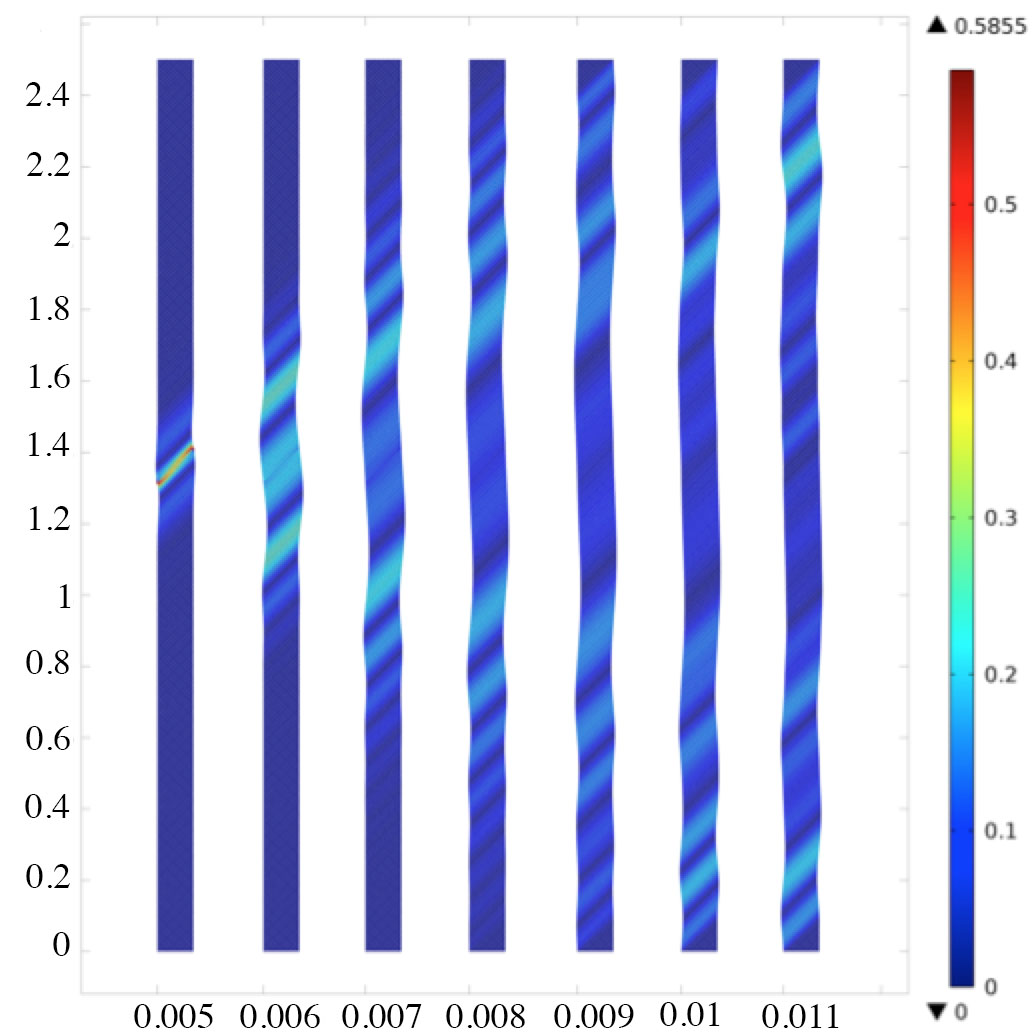}
\caption{Wave propagating after double impulse}
\end{center}
\end{figure}

\subsection{Internal discontinuities}

In Fig. 4 ($h=2$ m), the structure is provided with an horizontal set of hinges at the middle height. In the plot the local bending moment is represented by means of a color map (in N\,m). The hinges, in this case, do interrupt the continuity of the beams, allowing energy-free angular displacements between the upper and lower part of each beam. As it is observable, however, due to the kind of internal connections between the whole system of beams, this do not change the overall character of wave propagation.

In Fig. 5 ($h=2$ m), the upper and lower half of the system are connected by an array of vertical beams ($k_M = 1.96 \times 10^{-2}$ N m$^2$ and $k_N = 7.85 \times 10^4$ N). In this case, the imposed displacement is parallel to one of the two families of beams, and what is interesting is that the energy of the system remains more or less confined in the upper half, and waves practically do not propagate beyond the discontinuity, which therefore results in a simplified but potentially useful model for damping filters in the considered kind of structures.  

\begin{figure}
\begin{center}
\includegraphics[width=0.6\textwidth]{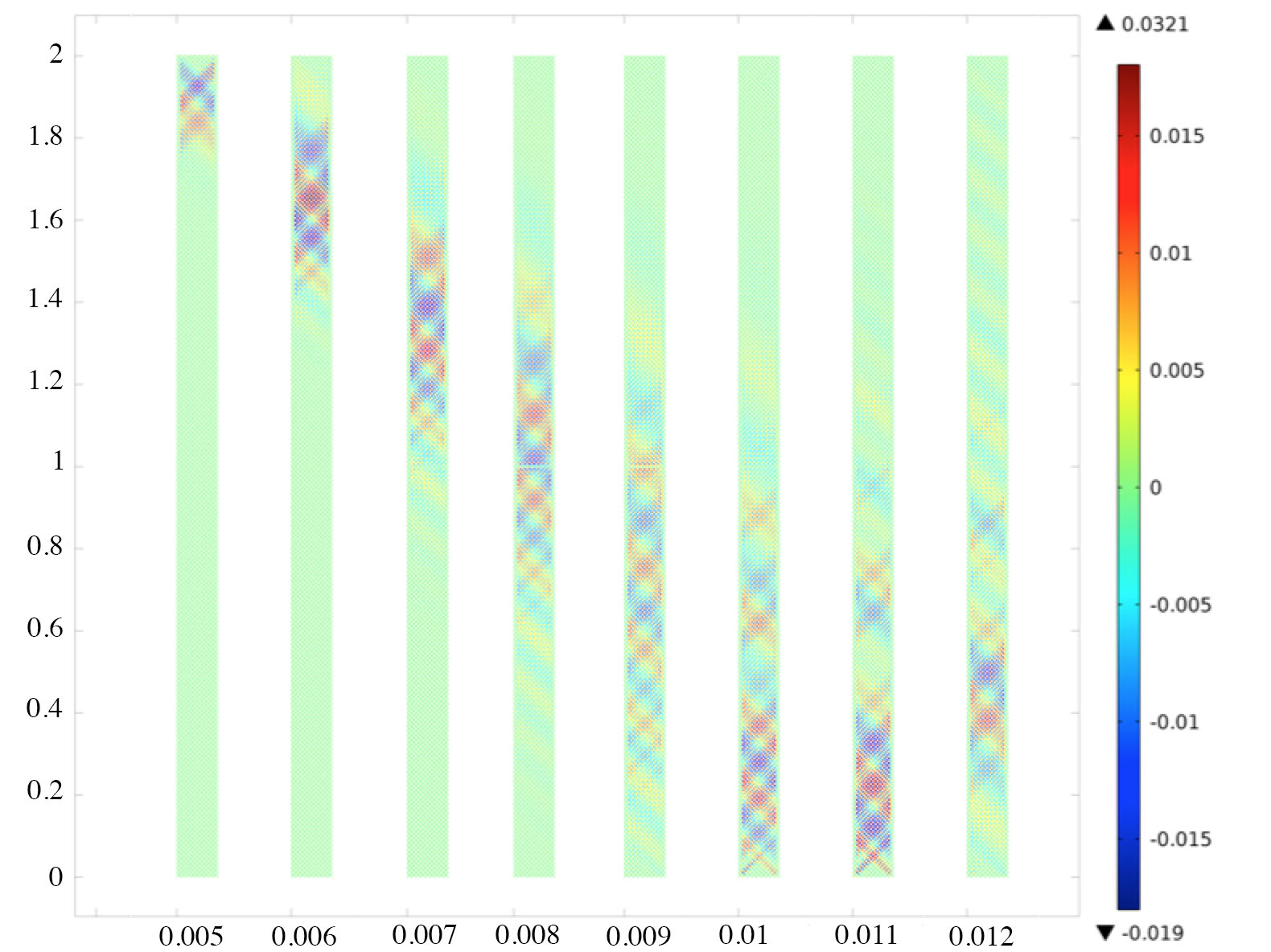}
\caption{Bending moment in two lattices connected by hinges}
\end{center}
\end{figure}

\begin{figure}
\begin{center}
\includegraphics[width=0.6\textwidth]{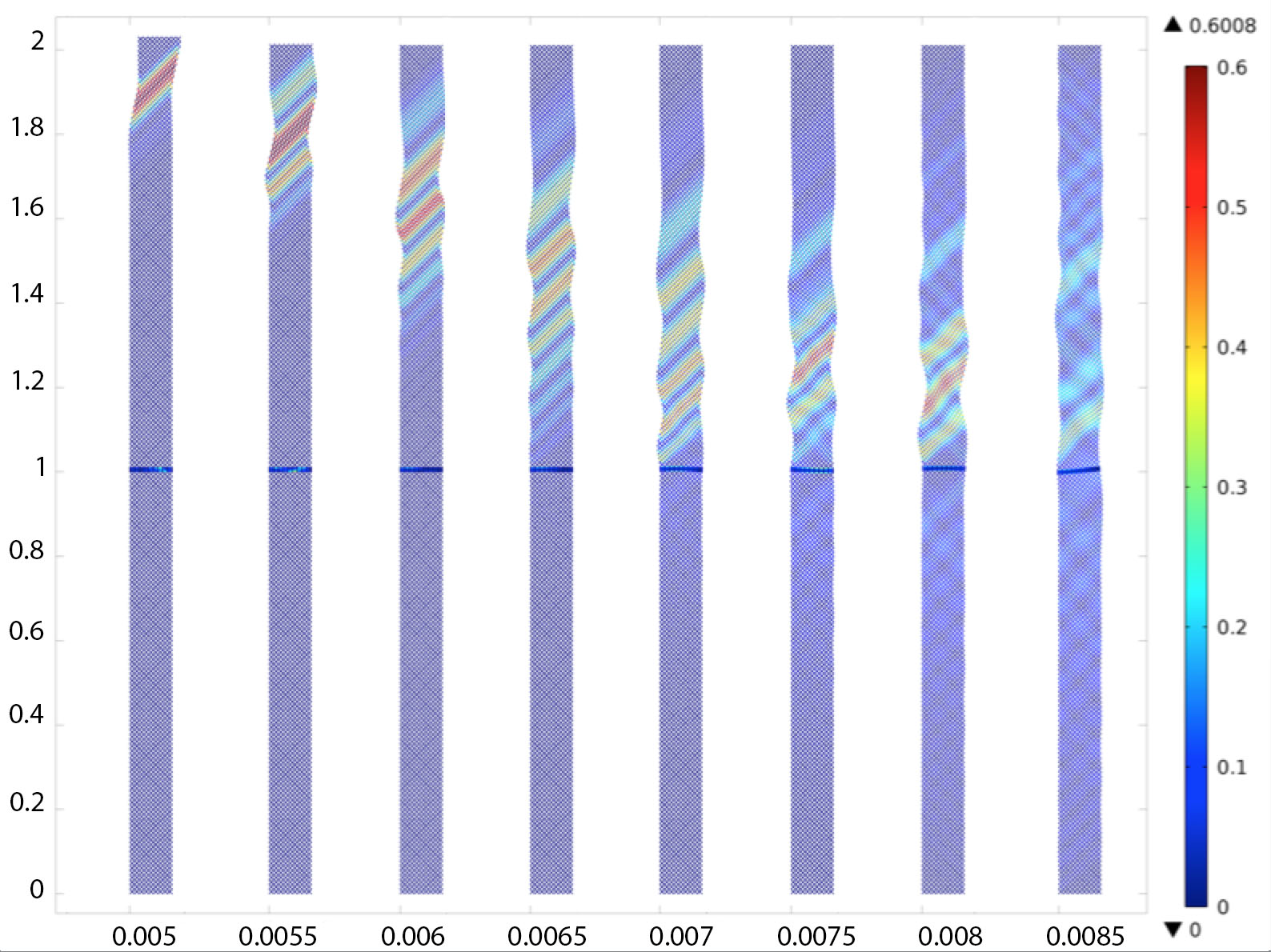}
\caption{Wave propagation in two lattices connected by an array of vertical beams}
\end{center}
\end{figure}

\subsection{Waves traveling in opposite directions}

Finally, in Fig. 6 ($h=2.5$ m), the initial impulse (parallel to one of the two families of beams) is imposed on both upper and lower sides. Two waves traveling in opposite directions originate and their interaction is shown. As one can see, the velocity of both wave fronts remains more or less unchanged after the crossing over. Moreover, in every snapshot a rather well-circumscribed traveling region displaying maximum perturbation is observable. These characteristics are shared by the well known self-supporting, localized traveling perturbations - called `solitons' - originating in particular non-linear systems. 

\begin{figure}
\begin{center}
\includegraphics[width=0.6\textwidth]{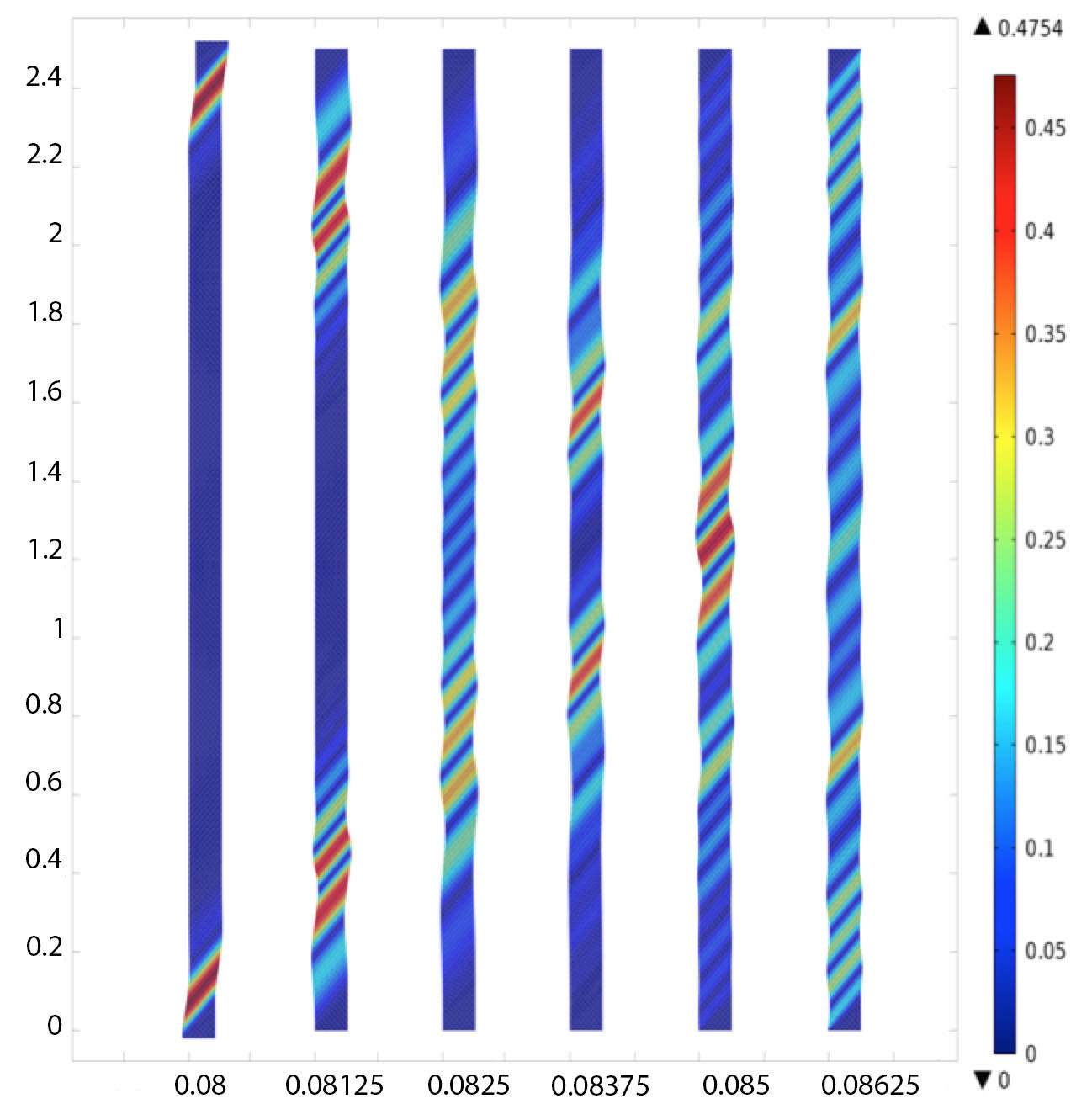}
\caption{Propagation of two waves traveling in opposite directions}
\end{center}
\end{figure}

\section{Conclusions}
A conjecture that can be validly proposed in this context is the possibility of emerging of solitons, i.e., solitary waves propagating without changing their shape and speed due to the balance of all physical effects (e.g. between dispersion and nonlinearity~\cite{Salupere2004}). Solitons were first observed in numerical simulations while studying the well-known Korteweg - de Vries (KdV) equation,
\begin{equation}
u_t=Kuu_x-u_{xxx}
\end{equation}
first appeared in hydrodynamics problems. The solutions of equations of this kind can be decomposed in localized perturbations with well defined shape which propagate at different velocities and preserve shape and velocity when interacting with other waves. A short but very clear historical review on solitons in elastic solids is presented by Maugin~\cite{maugin2011solitons}.
Zabusky and Kruskal~\cite{Zabusky1965} have demonstrated the emergence of a train of solitons from a harmonic initial condition for a given dispersion constant in the case of the KdV equation. The KdV equation is solved numerically using the pseudospectral  method (see~\cite{Fornberg1998,Salupere2009} for details). Soliton solutions are also appearing in nonlinear Cosserat models with a special coupling between rotations and deformations~\cite{Boehmer2015}.

In fact, the study of the propagation of a pulse along the fabric can be made using a discrete model consisting of two linear orders of one-dimensional beams of identical geometrical and mechanical characteristics, interacting each other through constraints which impose the continuity of the displacement in a finite number of pivot points. The intensity of the coupling can be adjusted by varying the value of the elastic constants and the mass density of the beams. In a one dimensional context, a beam discretized by means of concentrated masses and linear springs is studied e.g. in~\cite{Tarantino2010}. A case with dispersion is obtained by suitably adjusting the parameters that characterize the system. If we reduce the pulse duration and increase the value of the coupling constant, what happens is that the width of the wave packet (train) begins to be comparable with the pitch of the two rows of beams. The dispersion effects, as observed, are already visible in the simulations. Furthermore, the structure exhibits large rotations and hence the problem can be modeled taking into account the deformation of the structure when formulating the dynamic equations. On the whole, the structure and the model appear to be rich enough to allow the onset of true solitons if suitable non-linearities are considered. The soliton-like character of the perturbations shown in Fig. 6 suggest that further investigation in this direction would indeed be very interesting.


\end{document}